\documentclass{article}
\usepackage[preprint]{spconf}
\usepackage{amsmath,graphicx}

\usepackage{cite} 
\usepackage{appendix}
\usepackage{xcolor}
\usepackage{amsmath,amssymb,amsfonts}
\usepackage{booktabs}
\usepackage{multicol,multirow}
\usepackage{array}
\newcolumntype{M}[1]{>{\centering\arraybackslash}m{#1}}
\copyrightnotice{\copyright\ IEEE}
\toappear{To appear in {\it Proc.\ ICASSP 2023}}

\title{TriAAN-VC: Triple Adaptive Attention Normalization \\ for Any-to-Any Voice Conversion}

\name{Hyun Joon Park$^\ast$, Seok Woo Yang$^\ast$, Jin Sob Kim, Wooseok Shin, and Sung Won Han$^\ast$$^\ast$\thanks{$^\ast$ Equal contribution.}\thanks{$^\ast$$^\ast$ Corresponding author.}\thanks{This research was supported by a Korea TechnoComplex Foundation Grant (R2112651, R2112652) and Korea University Grant (K2107521, K2202151). This research was also supported by Brain Korea 21 FOUR.}}
\address{School of Industrial and Management Engineering, Korea University, Seoul, Republic of Korea}

\begin{document}
  
%
\maketitle
%

\begin{abstract}
Voice Conversion (VC) must be achieved while maintaining the content of the source speech and representing the characteristics of the target speaker. The existing methods do not simultaneously satisfy the above two aspects of VC, and their conversion outputs suffer from a trade-off problem between maintaining source contents and target characteristics. In this study, we propose Triple Adaptive Attention Normalization VC (TriAAN-VC), comprising an encoder-decoder and an attention-based adaptive normalization block, that can be applied to non-parallel any-to-any VC. The proposed adaptive normalization block extracts target speaker representations and achieves conversion while minimizing the loss of the source content with siamese loss. We evaluated TriAAN-VC on the VCTK dataset in terms of the maintenance of the source content and target speaker similarity. Experimental results for one-shot VC suggest that TriAAN-VC achieves state-of-the-art performance while mitigating the trade-off problem encountered in the existing VC methods.

\end{abstract}

\section{Introduction}
\label{sec:intro}

VC is the task of transforming the voice of the source speaker into that of the target speaker while maintaining the linguistic content of the source speech. Traditional methods require parallel data for training VC models \cite{toda2007voice, desai2009voice} or cannot convert using unseen speakers \cite{kameoka2018stargan, kaneko2018cyclegan}. For the diverse utilization of VC, researchers have recently studied any-to-any (A2A) and one-shot VC, which can be applied to unseen speakers and require only one utterance of source and target speakers \cite{qian2019autovc, wu2020one, chou2019one, wu2020vqvc+, chen2021again, lin2021s2vc, gu2021mediumvc, lin2021fragmentvc, wang2022drvc}. To perform the conversion, they disentangle the utterances into content and speaker representations.

As vector quantization methods, \cite{wu2020one, wu2020vqvc+} utilized discrete codes as content and the difference between discrete and continuous features as speaker representations. However, representing content with discrete codes reduces time relationships, damaging content information. For attention-based conversion methods \cite{lin2021fragmentvc, lin2021s2vc}, \cite{lin2021s2vc} suggested self-supervised learning features can improve VC performance. Although their results were highly similar to the target speaker characteristics, the conversion method using overly detailed speaker representation biased the results only to speaker similarity. Inspired by image style transfer, \cite{chou2019one, chen2021again, gu2021mediumvc} adopted Adaptive Instance Normalization (AdaIN) \cite{huang2017arbitrary} for conversion. \cite{chou2019one} used only high-level speaker features for AdaIN, causing results to be biased to speaker similarity. \cite{chen2021again} alleviated the problem by exploiting multi-level target speaker features for AdaIN, but AdaIN cannot represent enough speaker characteristics.

Although the previous methods achieved significant improvements in A2A VC, their methods utilized overly detailed or generalized speaker representations; therefore, the conversion results satisfied only one aspect of VC (i.e., maintenance of source content or similarity to the target speaker). This underscores the necessity for a conversion method using core speaker representations to mitigate the trade-off problem.

We propose Triple Adaptive Attention Normalization VC (TriAAN-VC) for non-parallel A2A VC. TriAAN-VC, which is based on an encoder-decoder structure, disentangles content and speaker features. TriAAN block extracts each detailed and global speaker representation from disentangled features and uses adaptive normalization for conversion. As a training approach, siamese loss with time masking is applied to maximize the maintenance of the source content. In A2A one-shot VC, a comparison of results on the VCTK dataset shows that TriAAN-VC achieves state-of-the-art performance in terms of both evaluation metrics, namely, maintenance of source content and similarity to the target speaker.

\begin{figure*}[htp]
\centerline{\includegraphics[scale = 0.53]{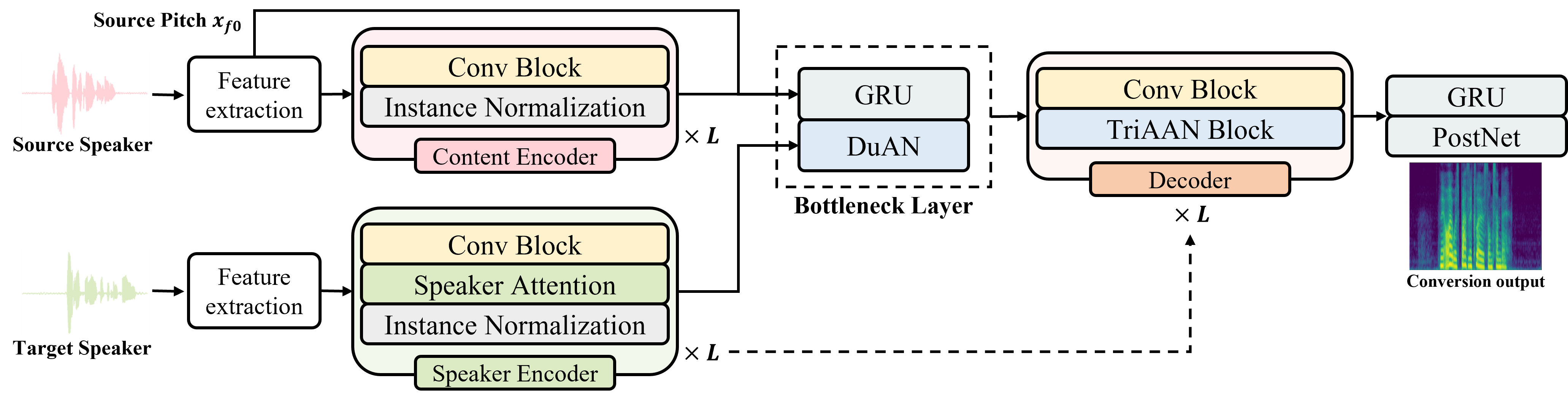}}
\caption{Overall architecture of TriAAN-VC.}
\label{fig:triaan_vc}
\end{figure*}

\section{Method}
\label{sec:method}

\subsection{Feature extraction}
\label{sec:feature extraction}

As \cite{lin2021s2vc} suggested Contrastive Predictive Coding (CPC) features \cite{oord2018representation} contribute to the improvement of VC performance, we adopt CPC features as inputs for the model. We use a pre-trained model \cite{riviere2020unsupervised} to extract CPC features $x \in \mathbb{R}^{H \times T}$ from raw audio $x_{raw} \in \mathbb{R}^{t}$, where $H$ and $T$ are the hidden size and segment length of $x$ and $t$ is the signal length of $x_{raw}$. Furthermore, to represent the pitch information of the source speaker, the log fundamental frequency (f0) $x_{f0} \in \mathbb{R}^{T}$ is extracted by applying the DIO algorithm to $x_{raw}$, as in \cite{wang2021vqmivc}.

\subsection{Encoder and Decoder}
\label{sec:encoder and decoder}

As shown in Figure \ref{fig:triaan_vc}, TriAAN-VC comprises two encoders, extracting content and speaker information respectively, and a decoder. The encoders and decoder are connected via a bottleneck layer, and each contains $L$ layers.

Before the encoders, we apply a convolution layer on $x_{c,s} \in \mathbb{R}_{c,s}^{H \times T}$ to expand $H$ to channel size $C$, where $x_{c,s}$ are the content and speaker inputs after feature extraction.

\noindent \textbf{Encoder.} Each encoder layer consists of a convolution block and Instance Normalization (IN). Speaker Attention (SA) is used only in the speaker encoder. The convolution block is designed as a residual block comprising two convolution layers with a kernel size of 3 and a stride of 1.

\noindent \textbf{Bottleneck layer.} After the encoder process, f0 of the source speaker $x_{f0} \in \mathbb{R}^{T}$ is used to represent the pitch. Given $x'_{c} \in \mathbb{R}^{C \times T}$ is the output of the content encoder, we apply a Gated Recurrent Unit (GRU) layer on the concatenated outputs between $x'_{c}$ and $x_{f0}$. Before the decoder, the initial converted representation is generated by applying Dual Adaptive Normalization (DuAN), a conversion method described in Section \ref{sec:triaan block}, to the content and speaker representations.

\noindent \textbf{Decoder.} Each decoder layer contains a convolution block, the same as that of encoders, and TriAAN block. TriAAN block conducts conversion using the content feature from the previous layer and gathered feature maps from the speaker encoder layers. Finally, the outputs of the decoder are refined by GRU layers and PostNet \cite{wang2017tacotron} to predict the log mel-spectrogram $\hat{y} \in \mathbb{R}^{M \times T}$, where $M$ is the number of mel bins.

\subsection{Speaker Attention}
\label{sec:speaker attention}

Since AdaIN conversion process utilizes channel-wise statistics of speaker representation, extracting core channel features of speaker is necessary. To achieve it, we modify IN as Time-wise IN (TIN) and design TIN-based Speaker Attention (SA). In contrast to IN, TIN normalizes with the time-wise mean and standard deviation, preserving channel relations. For SA, we utilize TIN and self-attention \cite{vaswani2017attention} as follows: 
%
\begin{equation}\label{eq:speaker attention}
  \begin{gathered}
    Q=\mathrm{TIN}(x_{s})W_{q},\ \ K=x_{s}W_{k},\ \ V=x_{s}W_{v} \\ 
    \mathrm{Attention}(Q,K,V)=\mathrm{softmax}(QK^\top/\sqrt{d})V
  \end{gathered}
\end{equation}
%
$x_{s}\in \mathbb{R}^{T \times C}$ and $W_{q,k,v} \in \mathbb{R}^{C \times C}$ denote speaker features and each weight. Using query information as TIN results, SA emphasizes and preserves the channel relations of speaker features used as speaker information for conversion.

\begin{figure}[h]
\centerline{\includegraphics[scale = 0.6]{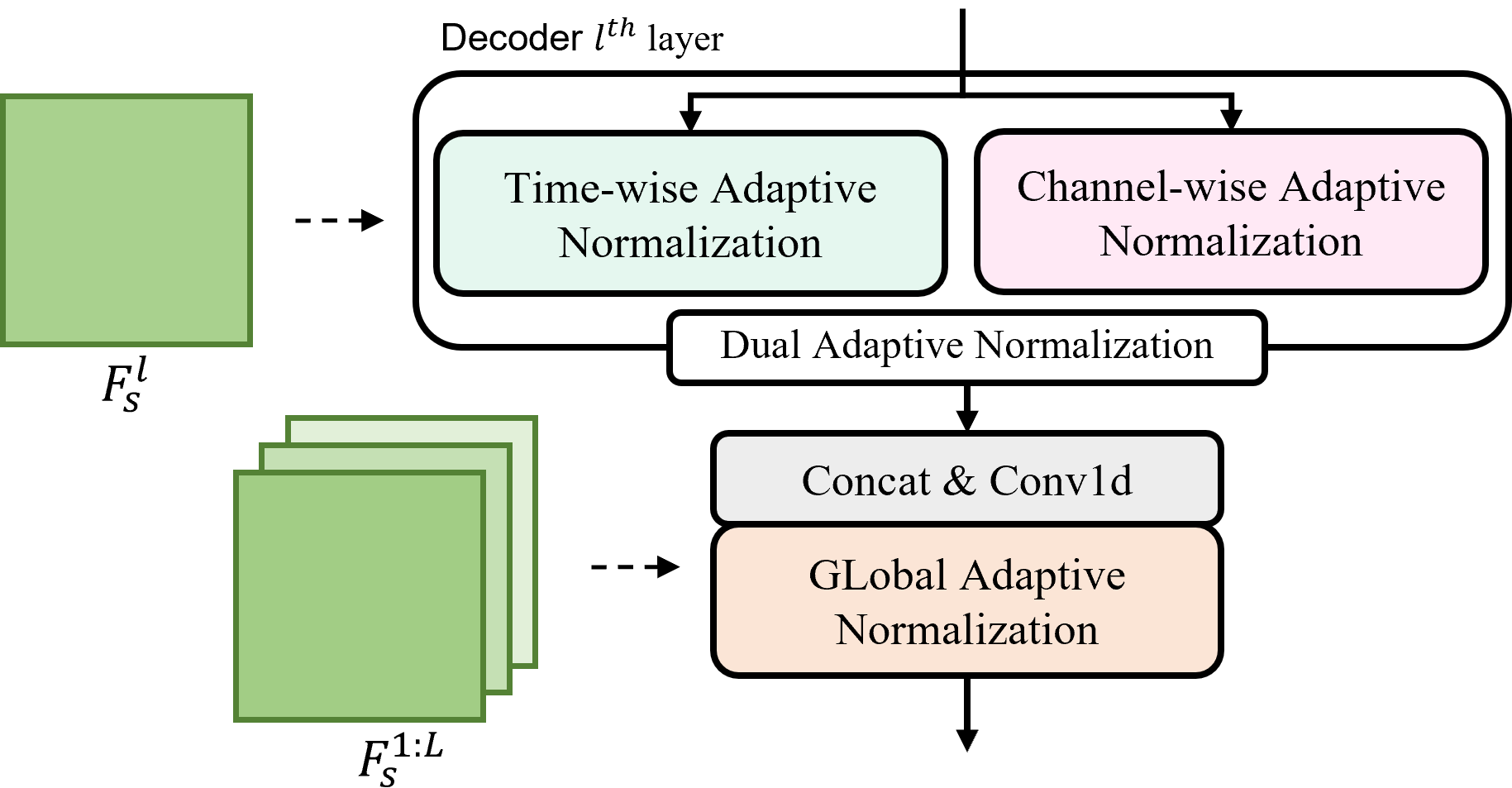}}
\caption{TriAAN Block}
\label{fig:triaan_block}
\end{figure}

\subsection{TriAAN block}
\label{sec:triaan block}

As depicted in Figure \ref{fig:triaan_block}, we design TriAAN block consisting of Dual Adaptive Normalization (DuAN) and GLobal Adaptive Normalization (GLAN) for the conversion process. TriAAN block uses gathered feature maps $F_{s}^{1:L} \in \mathbb{R}_{1:L}^{T \times C}$ from each $l^{th}$ speaker encoder layer, where $l={1,2,...,L}$. DuAN extracts layer-wise detailed speaker features from $F_{s}^{l}$ in dual view (i.e., time and channel) and performs adaptive normalization. By contrast, GLAN uses all feature maps from the speaker encoder to extract global speaker information.

\noindent \textbf{DuAN.} Inspired by adaptive attention normalization in image style transfer \cite{liu2021adaattn}, we design DuAN to extract detailed speaker features and to conduct conversion. DuAN represents attention-based statistics of layer-wise speaker features $F_{s}^{l}$. Given $x_{c} \in \mathbb{R}^{T \times C}$ is the content feature from the previous layer and $\mathcal{N}(\ \cdot \ )$ is the normalization function, the attention weight $\alpha \in \mathbb{R}^{T \times T}$, attention-weighted mean $M\in \mathbb{R}^{T \times C}$, and variance $Var\in \mathbb{R}^{T \times C}$ are defined as follows:
%
\begin{equation}\label{eq:duaan}
  \begin{gathered}
    Q=\mathcal{N}(x_{c})W_{q},\ \ K=\mathcal{N}(F_{s}^{l})W_{k},\ \ V=F_{s}^{l}W_{v} \\ 
    \alpha=\mathrm{softmax}(QK^\top/\sqrt{d}) \\
    M=\alpha V,\ \ Var=\alpha(V \cdot V) - M \cdot M
  \end{gathered}
\end{equation}
%
\noindent $W_{q,k,v} \in \mathbb{R}^{C \times C}$ denotes each weight for linear transformation. $\alpha$, obtained by the normalized feature, represents the similarity between the content and speaker features. Furthermore, $Var$ is calculated using the expectation of variables and the square of the variable expectations. By applying $\alpha$, the weighted mean and variance contain detailed speaker features, that is per-point statistics. To prevent biased results with excessively detailed speaker features, we take the time-wise average of $M$ and $Var$, followed by applying a square root on $Var$ to obtain the standard deviation $S \in \mathbb{R}^{C}$. The converted representation $x'_{c} \in \mathbb{R}^{T \times C}$, obtained by adaptive normalization and the content feature, is defined as $x'_{c}=\mathrm{IN}(x_{c})S +M$. To perform it in terms of channel and time, we separate the adaptive normalization process depending on $\mathcal{N}(\ \cdot \ )$ function (i.e., IN and TIN), making two converted representations.

\noindent \textbf{GLAN.} To represent the global speaker information, we utilize all feature maps $F_{s}^{1:L}$ from the speaker encoder. For the content feature $x_{c}\in \mathbb{R}^{C \times T}$ in GLAN, we apply a convolution layer to the channel-wise concatenation of two converted representations from DuAN. We obtain layer-wise concatenated means $\mu \in \mathbb{R}^{L \times C}$ and standard deviations $\sigma \in \mathbb{R}^{L \times C}$ from $F_{s}^{1:L}$, defined as $\mu=[\mathrm{avg}(F_{s}^{1});...,;\mathrm{avg}(F_{s}^{L})]$ and $\sigma=[\mathrm{std}(F_{s}^{1});...,;\mathrm{std}(F_{s}^{L})]$. To extract 
core statistics from global speaker features (i.e., $\mu$ and $\sigma$) for adaptive normalization, we adopt self-attention pooling \cite{cai2018exploring}, which emphasizes important speaker statistics. The attention pooling process used to obtain the weighted mean $\mu' \in \mathbb{R}^{C}$ and standard deviation $\sigma' \in \mathbb{R}^{C}$ is as follows: 
%
\begin{equation}\label{eq:glan}
  \begin{gathered}
    \alpha_{\mu} = \mathrm{softmax}(\mu W_{\mu}),\ \ \alpha_{\sigma} = \mathrm{softmax}(\sigma W_{\sigma}) \\
    \mu' = \mathrm{sum}(\mu \times \alpha_{\mu}),\ \ \sigma' = \mathrm{sum}(\sigma \times \alpha_{\sigma})
  \end{gathered}
\end{equation}
%
$\alpha_{\mu,\sigma} \in \mathbb{R}_{\mu,\sigma}^{L}$ and $W_{\mu,\sigma} \in \mathbb{R}_{\mu,\sigma}^{C \times C}$ denote attention weights and each weight for transformation. Then, adaptive normalization as in DuAN is applied with $\mu'$ and $\sigma'$ for conversion.

\begin{table*}
\caption{Objective evaluation results on the VCTK dataset for any-to-any one-shot voice conversion which uses one target utterance per sample. SV indicates the acceptance rate over the threshold determined by the equal error rate in the dataset.}\label{tab:main}
\centering
\resizebox{1.9\columnwidth}{!}{%
\begin{tabular}{p{2.5cm}M{1cm}M{1cm}M{1cm}cM{1.2cm}M{1.2cm}M{1.2cm}cM{1.1cm}M{1.1cm}M{1.1cm}}
\toprule
\multirow{2.5}{*}{Model} & \multicolumn{3}{c}{Word Error Rate (WER \%)} & & \multicolumn{3}{c}{Character Error Rate (CER \%)} & & \multicolumn{3}{c}{Speaker Verification (SV \%)} \\ 
\cmidrule{2-4} \cmidrule{6-8} \cmidrule{10-12} & S2S & U2U & Avg. & & S2S & U2U & Avg. & & S2S & U2U & Avg. \\
\midrule
AUTO-VC \cite{qian2019autovc} & 29.96 & 27.33 & 28.64 & & 15.98 & 14.75 & 15.36 & & 36.00 & 20.84 & 28.42 \\
AdaIN-VC \cite{chou2019one}   & 47.33 & 46.84 & 47.08 & & 28.16 & 27.79 & 27.98 & & 89.34 & 81.17 & 85.25 \\
AGAIN-VC \cite{chen2021again} & 28.89 & 26.45 & 27.67 & & 15.65 & 14.52 & 15.08 & & 71.33 & 67.00 & 69.17 \\
VQVC+ \cite{wu2020vqvc+}      & 52.92 & 53.02 & 52.97 & & 30.57 & 31.69 & 31.13 & & 76.34 & 55.33 & 65.83 \\
S2VC \cite{lin2021s2vc}       & 44.99 & 42.65 & 43.82 & & 25.88 & 24.89 & 25.38 & & 95.34 & 89.17 & 92.25 \\
VQMIVC \cite{wang2021vqmivc}  & 29.30 & 28.05 & 28.67 & & 15.71 & 15.04 & 15.37 & & 86.33 & 35.67 & 61.00 \\
\midrule
\textbf{TriAAN-VC} & \textbf{20.73} & \textbf{22.35} & \textbf{21.54} & & \textbf{10.79} & \textbf{11.69} & \textbf{11.24} & & \textbf{96.00} & \textbf{89.67} & \textbf{92.83} \\
\bottomrule
\end{tabular}}
\end{table*}

\subsection{Loss function}
\label{sec:loss function}

To train TriAAN-VC, we combine reconstruction loss and siamese loss. Reconstruction loss is $L1$ loss between the ground truth mel-spectrogram  $y \in \mathbb{R}^{M \times T}$ and predicted mel-spectrogram $\hat{y} \in \mathbb{R}^{M \times T}$. $y$ is extracted from the raw audio using a mel-spectrogram transformation, and $\hat{y}$ is predicted by the proposed model when input features $x \in \mathbb{R}^{H \times T}$ are CPC features of the raw audio. For siamese loss, $L1$ loss is applied between $y$ and $\hat{y}_{siam}\in \mathbb{R}^{M \times T}$, where $\hat{y}_{siam}$ is predicted by the model with $x$ augmented by time masking. 
Given $L1$ loss is $loss(y,\hat{y})=||y-\hat{y}||_{1}/T$, the combined loss is as follows:
\begin{equation}
\begin{gathered}
\label{eq:loss}
\resizebox{.91\hsize}{!}{$\mathcal{L}=(loss(y, \hat{y}) + loss(y,\hat{y}_{siam}))/2 + loss(\hat{y}, \hat{y}_{siam})$} \\
\end{gathered}
\end{equation}
\noindent By calculating the additional loss with $\hat{y}_{siam}$, the robustness and consistency of the model can be improved. In particular, since time masking removes content information during training, the loss with the siamese branch makes the model robust for maintaining content information.

\section{Experiments}
\label{sec:exp}

\subsection{Experimental setup}
\label{sec:setup}

For comparison, we use the VCTK dataset \cite{veaux2016superseded} containing about 400 utterances per 109 speakers. We split the dataset into ratios of 60\%, 20\%, and 20\% for train, validation, and test set, respectively, considering speakers and utterances. For conversion scenarios, we select 20 speakers and generate 600 pairs per Seen-to-Seen (S2S) and Unseen-to-Unseen (U2U) scenarios. After downsampling the audio to 16 kHz, we extract CPC, f0, and log mel-spectrogram features based on a frame size of 25ms, hop size of 10ms, and mel bins of 80. 

For training details, we use a batch size of 64, an epoch of 400, and Adam optimizer with a learning rate of $10^{-5}$. For model parameters, we take $H=256, C=512$, and $L=6$. We adopt a Parallel WaveGAN vocoder \cite{yamamoto2020parallel} pre-trained on the VCTK dataset to convert log mel-spectrograms to waveforms. For comparison, benchmark models are reproduced using their official codes. They are trained with mel-spectrogram features except for S2VC which uses CPC features. The conversion results are available on the demo page.\footnote{https://winddori2002.github.io/vc-demo.github.io/}

\subsection{Evaluation metrics}
\label{sec:metric}

We adopt objective and subjective measures for evaluation. For objective measures, the models are evaluated in respect of two aspects of VC (i.e, maintenance of source content and similarity to the target speaker). Word Error Rate (WER) and Character Error Rate (CER) are used to evaluate the error rate of scripts between the source and converted utterances. The script of the converted utterances is extracted by a pre-trained Wav2Vec 2.0 \cite{baevski2020wav2vec}. For speaker similarity, we adopt the acceptance rate based on Speaker Verification (SV) model as in \cite{lin2021fragmentvc,lin2021s2vc}. The score is measured by the cosine similarity between the target and converted embedding vectors, extracted by the SV model, and the threshold which is determined based on the equal error rate in the VCTK dataset. 

In the subjective evaluation, we conduct Mean Opinion Score (MOS) test for naturalness and similarity. Subjects are asked to assign a score from 1 to 5 after listening to converted utterances or a pair of target and converted utterances for naturalness and similarity evaluation. We perform a test on 15 subjects using randomly selected 20 pairs of utterances for the S2S and U2U scenarios, respectively.
\begin{figure}[ht]
\centerline{\includegraphics[scale = 0.58]{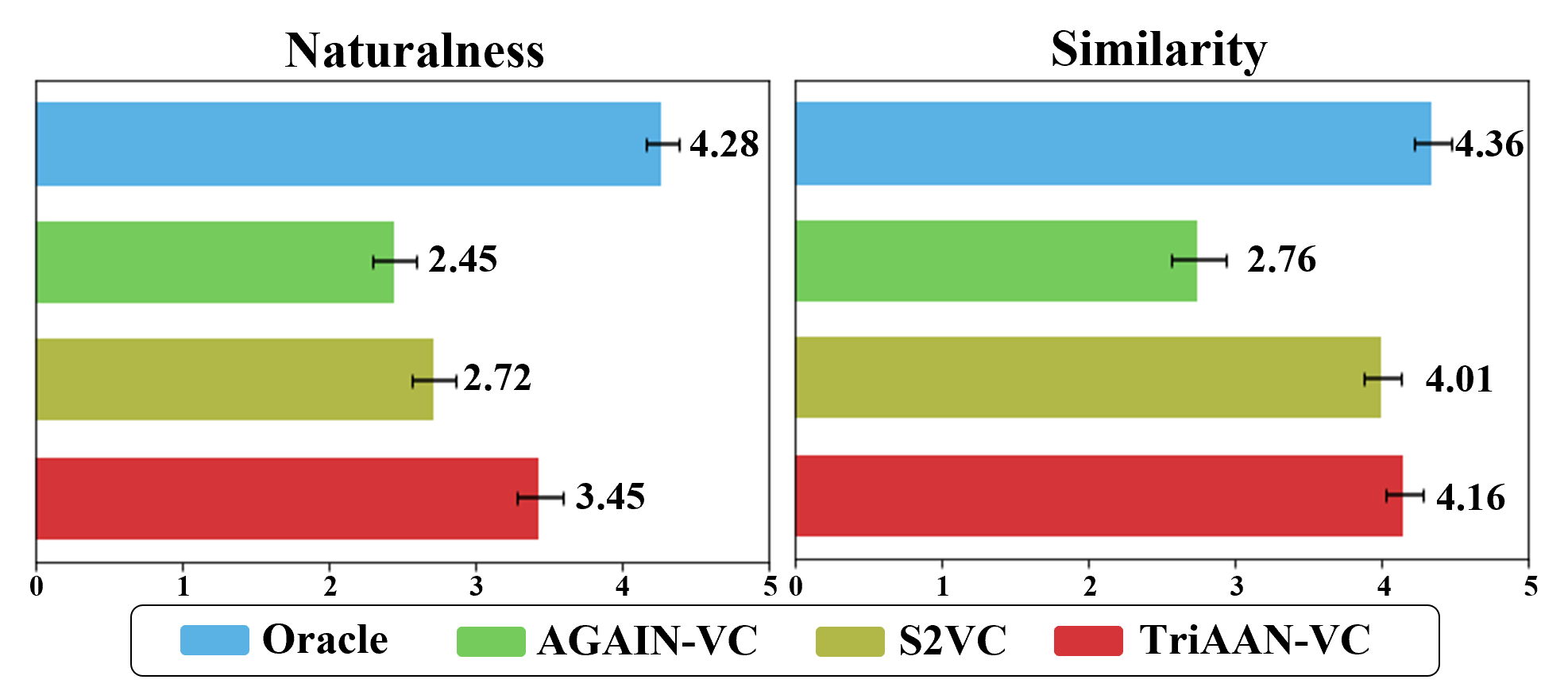}}
\caption{MOS results with 95\% confidence intervals for naturalness and similarity.}
\label{fig:mos results}
\end{figure}
\subsection{Experimental results}
\label{sec:experiment results}

\textbf{Comparison results.} We conducted the experiment for one-shot VC to compare the proposed model with the previous methods using objective and subjective measures. As indicated in Table \ref{tab:main}, TriAAN-VC achieved better performance on WER, CER, and SV scores, regardless of conversion scenarios compared to the existing methods which suffered from a trade-off problem of VC. It suggests that the conversion methods using compact speaker features can simultaneously retain both source content and target speaker characteristics.

Figure \ref{fig:mos results} depicts the average MOS results of the S2S and U2U scenarios, and it includes oracles reconstructed by the vocoder. Similar to the results of objective evaluation, TriAAN-VC demonstrated a slight improvement over S2VC in terms of similarity, which is close to the performance of the oracle. Furthermore, TriAAN-VC outperformed the previous methods in terms of naturalness evaluation, suggesting the proposed model can make relatively unbiased results.


\noindent \textbf{Further experiment.} As ablation studies, we analyze the contributions of the proposed components. As listed in rows 1-3 of Table \ref{tab:ablation}, the use of each CPC feature and siamese loss contributed significantly to the performance gain of WER and SV. In rows 4-6 of Table \ref{tab:ablation}, we excluded one of the components of TriAAN-VC without siamese loss. The results suggested that SA particularly contributed to the improvement of WER, and TriAAN block was the crucial component for the performance gain of SV. Although each component contributed to the performance gain in WER or SV, they also suffered from the trade-off problem, implying all the components are necessary to mitigate the trade-off problem. In addition to one-shot VC, TriAAN-VC was effective in multi-utterance scenarios. Under the multiple utterance setting using more than one target utterance, TriAAN-VC with CPC improved its performance by about 4\% and 5\% on WER and SV, compared to the one-shot VC results.

\begin{table}[th]
\caption{Results of ablation studies and multi-utterance scenarios. ${\dagger}$ indicates TriAAN-VC without siamese loss.}
\label{tab:ablation}
\centering
\resizebox{\linewidth}{!}{%
\begin{tabular}{p{3cm}|M{1.2cm}M{1.2cm}M{1.2cm}M{1.2cm}}
\toprule
\multirow{2.5}{*}{Model} & \multicolumn{2}{c}{WER \%} & \multicolumn{2}{c}{SV \%} \\ 
\cmidrule{2-3} \cmidrule{4-5} & S2S & U2U   & S2S & U2U \\

\midrule
TriAAN-VC \text{+} Mel              & 27.31 & 27.07 & 90.34 & 88.34 \\
TriAAN-VC \text{+} CPC              & 20.73 & 22.35 & 96.00 & 89.67 \\ 
TriAAN-VC$^{\dagger}$ \text{+} CPC  & 24.85 & 26.37 & 94.67 & 89.67 \\ 

\hspace{5pt} \text{-} SA            & 28.16 & 29.87 & 93.84 & 90.17 \\ 
\hspace{5pt} \text{-} GLAN          & 20.88 & 21.11 & 92.50 & 87.17 \\
\hspace{5pt} \text{-} DuAAN         & 20.58 & 20.93 & 90.00 & 83.00 \\

\midrule
3-utterance scenario & 16.73 & 18.45 & 99.00 & 96.33 \\
5-utterance scenario & 16.90 & 17.82 & 98.50 & 98.17 \\

\bottomrule
\end{tabular}}
\end{table}

\section{Conclusion}
\label{sec:con}
In this study, we proposed TriAAN-VC for non-parallel A2A VC, which extracts compact speaker features and performs adaptive normalization for conversion. The results of A2A VC on the VCTK dataset indicate that TriAAN-VC achieves outstanding performance, including in multi-utterance scenarios. Unlike previous methods that suffer from a trade-off in VC, TriAAN-VC with siamese loss satisfies two aspects of VC. Finally, ablation studies suggest the necessity of all proposed methods to mitigate the trade-off problem of VC.

\bibliographystyle{IEEEbib}
\bibliography{paper}

\end{document}